# Doping a semiconductor to create an unconventional metal

N. Manyala[1,3], J. F. DiTusa[1], G. Aeppli[2] and A.P. Ramirez[4]

[1]*Department of Physics and Astronomy, Louisiana State University, Baton Rouge LA 70803,*  [2]*London Centre for Nanotechnology and Department of Physics and Astronomy, UCL, London WC1E 6BT, UK,* [3]*Department of Physics and Electronics, National University of Lesotho, P. O. Roma 180, Maseru 100, Lesotho,* [4]*Bell Laboratories, Alcatel-Lucent, 600 Mountain Ave., Murray Hill, NJ, 07974*

(13 May 2008)

**Landau Fermi liquid theory, with its pivotal assertion that electrons in metals can be simply understood as independent particles with effective masses replacing the free electron mass, has been astonishingly successful. This is true despite the Coulomb interactions an electron experiences from the host crystal lattice, its defects, and the other $\sim 10^{22}/cm^3$ electrons. An important extension to the theory accounts for the behaviour of doped semiconductors[1,2]. Because little in the vast literature on materials contradicts Fermi liquid theory and its extensions, exceptions have attracted great attention, and they include the high temperature superconductors[3], silicon-based field effect transistors which host two-dimensional metals[4], and certain rare earth compounds at the threshold of magnetism[5-8]. The origin of the non-Fermi liquid behaviour in all of these systems**

**remains controversial. Here we report that an entirely different and exceedingly simple class of materials – doped small gap semiconductors near a metal-insulator transition - can also display a non-Fermi liquid state. Remarkably, a modest magnetic field functions as a switch which restores the ordinary disordered Fermi liquid. Our data suggest that we have finally found a physical realization of the only mathematically rigourous route to a non-Fermi liquid, namely the 'undercompensated Kondo effect', where there are too few mobile electrons to compensate for the spins of unpaired electrons localized on impurity atoms[9-12].**

Combining two of the most common elements – iron and silicon - in an electric arc furnace yields FeSi, the much studied cubic semiconductor, with a narrow (15 THz) gap between filled valence and empty conduction bands[13-15]. In contrast to Si, for which the solubility of substitutional impurities is highly limited, FeSi forms isostructural dilution series with metallic MnSi and CoSi, as well as accepting substitution of dopants such as Al for up to 30% of Si[16-21]. We focus here on Mn doping onto the Fe site, which earlier work has shown to yield one missing electron (hole) per Mn atom, as predicted based on the naïve electron counting arguments, derived from relative valences deductible from the periodic table and applicable to the conventional doped semiconductors[22] which form the bedrock of modern microelectronics. Previous work has addressed hole doping via substitution of Al onto the Si site and electron doping via replacement of Fe by Co[16-21]. The two alloy series are quintessential exhibits of the success of Landau Fermi liquid theory with corrections for disorder[1,2].

Fig. 1a shows the zero field magnetic susceptibility $\chi = \delta M/\delta H$ of several samples where Fe has been replaced by Mn in FeSi. Despite the much studied[16,20, 21,23] ferromagnetism of pure MnSi and $Fe_{1-y}Co_ySi$ we find, just as in $FeSi_{1-z}Al_z$, no evidence of a magnetic phase transition down to 1.7K, the base temperature of the magnetometer, for $x<0.8$[17,19]. As does Al doping, Mn substitution induces both a temperature-independent Pauli contribution $\chi_{Pauli}$ from mobile (doping-induced) holes and a Curie-Weiss term ($C/(T-\Theta_W)$) from residual local moments (low temperature upturn)[17,19,22] added to the very large $\chi$, not accounted for by band theory, which appears for FeSi at temperatures much higher than those considered in the present experiments. However, the contribution from residual local moments is larger than for the Al-doped material[17,19] as we demonstrate in Fig. 1b and c, which show the associated Curie constant $C$ and Weiss temperature $\Theta_W$, measuring the degree to which the local moments can be aligned (for $\Theta_W>0$) to eventually form a ferromagnet, or are screened (for $\Theta_W<0$) by mobile charge carriers and/or by each other on account of antiferromagnetic interactions. The solid and dashed lines in Fig. 1b represent the Curie constants $C=g^2S(S+1)n$ for $g=2$ and $S=1/2$, $S=1$, and $S=3/2$ moments where we have fixed the number of magnetic moments, $n$, equal to the Mn (or Co, Al) dopant densities. For Mn, $\Theta_W$ (Fig. 1c) is less than zero corresponding to screening that increases with dopant density. The inset shows how $\Theta_W$ changes sign at large $x$ consistent with the ferromagnetism at the MnSi end of the alloy series. For Al, the results are consistent with a picture where each acceptor is associated with an $S=1/2$ impurity, and that this impurity is strongly screened as $T\rightarrow 0$. In contrast, for Mn, the moments evolve from having values close to $S=1$ and only reach $S=1/2$ well beyond the IM transition, and these larger moments are screened much less successfully. Although the Mn dopants act as acceptors, meaning that their electronic configuration should resemble that of Fe in FeSi having 6 $d$-electrons, we find that Mn ions prefer a high spin state which appears in nominally pure FeSi only above room temperature. Co dopants in FeSi are associated with an even larger spin ($S=3/2$), and an interaction $\Theta_W(x)$ which changes sign – to ferromagnetic - very close to the IM transition. Indeed, $\Theta_W(x)$ varies linearly with the electron count per transition metal ion (but not Al), with the pure host material, FeSi, being characterized by a very low AFM coupling.

Mobile carriers, responsible for the temperature-independent contribution to $\chi$, should also induce a dramatic increase in the electrical conductivity ($\sigma$). Fig. 2a displays the expected systematic increase of $\sigma$ with $x$ for several of our Mn-doped samples, while Fig. 1d shows the extrapolated (see below) zero temperature conductivities $\sigma_o$ as a function of Mn doping, $x$, and for comparison, previous results for Al doping, $z$, and Co doping, $y$[17-21]. While there are classic doping-induced insulator-metal (IM) transitions for all three impurities, the data exhibit significant differences between transition metal site, hole-doped, $Fe_{1-x}Mn_xSi$ and electron-doped $Fe_{1-y}Co_ySi$, and Si site, hole-doped $FeSi_{1-z}Al_z$. Although the carrier concentration at any particular $x$, $y$, or $z$ is indistinguishable for the three types of substitution[17,19-21], the low-$T$ $\sigma$'s of $Fe_{1-x}Mn_xSi$ and $Fe_{1-y}Co_ySi$ are two times smaller than those of the corresponding $FeSi_{1-z}Al_z$ samples (Fig. 1d and 2a)[17-21]. Furthermore, $\sigma$ of $Fe_{1-x}Mn_xSi$ below 20 K (Fig. 2a) displays a maximum which evolves systematically to lower $T$ with increasing doping. This insulating-like behavior is also apparent in $FeSi_{1-z}Al_z$ for $z < 0.01$[17], but persists for $Fe_{1-x}Mn_xSi$ up to $x = 0.1$. The result is a critical concentration for metallic behavior that is 3 to 4 times larger than in $FeSi_{1-z}Al_z$ and at least twice as large as in $Fe_{1-y}Co_ySi$[17-21].

Magnetotransport measurements provide detail about low energy electronic states and scattering in conducting solids, from doped semiconductors to rare earth intermetallics[1,2,5,20,24]. Generically, the low temperature conductivity in metals follows the form $\sigma = \sigma_0 + m_\sigma T^\alpha$. For an ordinary Fermi liquid where the dominant scattering is due to the Coulomb interaction between electrons, the asymptotic form is analytic, with a power law α=2; this result obtains even when isolated magnetic impurities are dissolved in a conventional metal because there are sufficient electrons to screen them[22]. On the other hand, near the insulator-to-metal (IM) transitions in carrier-doped semiconductors, nearly all previous investigations below 1 K have observed non-analytic behaviour with α = ½, due to a combination of quantum interference and electron-electron interactions in disordered Fermi liquids[1,2,17,19,20,24]. Because our $Fe_{1-x}Mn_xSi$ samples also span the IM transition, we expect to find the same result. As Fig. 2a reveals, our samples do have a significant $T$-dependent σ down to our lowest $T$'s. A quantitative analysis of the $T$-dependence is displayed in Fig. 2b where we plot α as a function of $x$ and $T$ determined by taking the logarithmic $T$-derivative of the σ data. α shows changes not commonly associated with an IM transition in more standard semiconductors[1,2,24]. Although the large values of α found on the insulating side of the IM transition, and the tendency for α to approach ½ at the lowest $T$'s well into the metallic regime, are standard, σ−σ₀ displays a nearly logarithmic $T$-dependence (i.e.

$\alpha \sim 0$) for $H=0$ and $x$ close to the critical concentration $x_c$ for the IM transition. While it is clear that $\alpha$ may not vanish rigorously as $T \to 0$, it becomes small and passes through an obvious minimum at $x_c$. This is in contrast to the $\alpha = \frac{1}{2}$ behavior found on both sides of the IM transition ($\sigma_0 = 0$ and $\sigma_0 > 0$) in $FeSi_{1-z}Al_z$, Si:P and other common semiconductors[17,20,24], although a small value for $\alpha$ (=0.22) has been reported near the Mott-Hubbard transition in $NiS_{2-x}Se_x$[25].

Our key discoveries are shown in Fig. 2c, and d, which display how this picture changes when an external field is applied to the samples. The changes to $\alpha$ with H (Fig. 2c) are most dramatic for $x \sim x_c$, where $H=9$ T effectively replaces the black $\alpha \sim 0$ column in the contour plot by extending the region of $x$ described by $\alpha \sim 0.5$. Fig. 2d shows the changes to $\alpha$ for the $x=0.02$ sample as a function of magnetic field and reveals that the field scale for this change is less than 1 T. Above 1 T, $\sigma(T)$ follows the standard form for disordered conductors with $\alpha=0.5$. From Fig. 2a, it is clear that the changes in $\alpha$ with $H$ are accompanied by increases in $\sigma_o$, the $T=0$ intercept of our fits. Most importantly, our $x = 0.01$ and $0.015$ samples, for which $\sigma_o(H=0) = 0$, display $\sigma_o = 12.5$ and $70$ $\Omega^{-1}cm^{-1}$ respectively for H=9 T. Thus, we observe a $T = 0$ insulator-to-metal transition with magnetic field in these samples. Fig. 1d, where the composition-dependent IM transition is shifted to $x_c(H)<0.01$ by the 9T external field, provides another view of the same phenomenon. The field-induced IM transition observed here is a new type of IM transition, between an insulator which in zero field has tendencies towards non-Fermi liquid behaviour, and a conventional disordered metal. This is in contrast to the manganites, which are driven by external magnetic fields towards their metallic ferromagnetic states, and $Gd_{3-x}S_4$, where the $x=0$ compound is a ferromagnetic metal and an external magnetic field drives an IM transition for x>0 by driving the Gd moments towards a more uniformly polarized 'ferromagnetic' state[26].

The magnetotransport data presented in Fig. 2 and the supplementary information, reveal that in contrast to FeSi doped with Al or Co, FeSi doped with Mn is a new non-Fermi liquid. A crucial clue as to the origin of this state is that very modest magnetic fields convert it into a conventional disordered Fermi liquid. The latter suggests a magnetic origin for the state, and accordingly, the susceptibility data, along with the higher field magnetization data also presented in the supplementary information, reveal that FeSi:Mn contains free or weakly screened spins at low $T$ which do not exist in FeSi:Al. If these moments were to remain truly unbound to the lowest temperatures, they should not be visible in the zero field entropy $S(T)$ for $T>0$, but become so on application of external fields. Because of the relation $\int_0^T C(T')/T' dT' = S(T)$, measurements of $C(T)/T$, where $C(T)$ is the specific heat, can be used to check this. The results, shown in Fig. 3a, closely resemble those for $FeSi_{1-z}Al_z$ and standard metals which are commonly characterized by a $C(T)/T = \gamma + \beta T^2$ dependence, where $\gamma$ represents the electronic contribution, also responsible for the finite $\chi_{Pauli}$, proportional to the effective hole mass, $m^*$, and $\beta T^2$ represents the lattice contribution[17,19,22]. The $x=0.015$ sample, which displays anomalous $\sigma(T)$ at $H=0$ in Fig. 2a, has a non-zero $\gamma$, as do our $FeSi_{1-z}Al_z$ samples at the same doping levels and Si:P just on the insulating side of the IM transition[17,19,27,28]. Extrapolation from $T$ above 5 K yields a moderately heavy $m^*$ which is enhanced by a factor of ~10 above the free electron mass, $m_e$, and 1.5+-0.5 $m_e$, the carrier (hole) mass determined from band structure calculations[15].

As $T$ is reduced below 5 K, $C(T)/T$ is further enhanced beyond the standard metallic form. For $x=0.015$, the zero-field result is describable by a power law $T^{-0.27+-0.02}$, while just beyond the IM transition, for $x=0.03$, there is a peak at 0.3 K. Increasing external magnetic fields produce peaks which move to larger temperatures, $T_{max}$. What is most important, though, is that the zero field entropy deduced from these finite temperature measurements is smaller than the in-field entropy. To make this clear, we present in Fig. 3b the difference between in-field and zero-field entropies, obtained from integrating the $C(T)/T$ data. This difference begins as a negative-going function with a minimum at a field-dependent cross-over temperature similar to $T_{max}(H)$, and then eventually exceeds zero. This violation of entropy balance can only be due to very low temperature (i.e. $T<50$mK) entropy being missed in the integration. The positive values measured for the entropy differences at $T=15$K correspond to a population of moments where roughly 5% of $S=1$ moments associated with each Mn atom remain unbound; this number is consistent with the more complex analysis of the magnetization (supplementary information), as should be the case given the Maxwell relations between $M(H,T)$ and $C(H,T)$.

The conclusion from all of the thermodynamic and magnetic measurements is that there are residual, unscreened spins in FeSi:Mn which are absent for FeSi:Al. The mechanism underlying the non-Fermi liquid behaviour seen in the magnetotransport data is therefore clear – we are dealing with an undercompensated Kondo problem[9-12] (Fig. 4a), where there are not enough free carriers, with spin $S=1/2$, to screen the remaining local moments associated with the acceptors. For this problem, down to the lowest temperatures, some impurities are always visible to the charge carriers, producing inelastic scattering of the carriers and thus destabilizing the Fermi liquid ground state. Fig. 4b-d illustrates the progression from high to low temperature, where initially (b), for $T>T_K\sim T_F>\Theta_W$, the moments are unbound, but subsequently, they first yield (c) the undercompensated Kondo non-Fermi liquid when $T_K>T>\Theta_W$. The next step (d) is for AFM correlations to appear among the local moments, thus reducing the total moments that the conduction electrons need to screen. If there is no magnetic phase transition, the eventual outcome will be a collective singlet state[29-31] among most local moments, with some isolated, unpaired local moments of the type routinely encountered in disordered quantum spin systems with predominantly AFM couplings. Because the latter still carry S=1 rather than S=1/2, the undercompensated Kondo effect still applies and there will be residual entropy and the associated non-Fermi liquid behaviour as T→0. External magnetic fields remove the low-lying degeneracies by locking the unpaired local moments, thus always returning the system to a disordered Fermi liquid.

We have discovered a new recipe for non-Fermi liquid behaviour, namely doping a narrow gap semiconductor with impurities which each donate a single hole to the valence band while retaining a net moment larger than S=1/2 which cannot be screened down to 50 mK. An important ingredient for our recipe is that the magnetic moments associated with the acceptors are not all lost via antiferromagnetic exchange interactions between themselves, as indeed they are in Si:P and FeSi:Al[17,19,27]. Dilute magnetic semiconductors, very popular for their potential in the growing area of spintronics and generally displaying conventional disordered metallic ferromagnetism, have thus been shown also to have a ground state of fundamental interest in the field of quantum many-body physics. For this reason as well as to determine whether any of the discoveries made here depend on a Kondo insulator[32] as the parent compound (we think not), it will be interesting to extend the current work to III-V semiconductors doped with magnetic transition metals, where it might also be possible to fabricate novel devices that examine, if not exploit, the unusual properties of the underscreened Kondo effect.

**Methods**: The $Fe_{1-x}Mn_xSi$, $Fe_{1-y}Co_ySi$, and $FeSi_{1-z}Al_z$ samples investigated in our experiments were all polycrystalline pellets produced from high purity (99.9985 %) starting materials by arc melting in an argon atmosphere. Samples were annealed for four days at 1000 $^0$C in evacuated quartz ampoules to improve homogeneity. We collected (Cu-K$\alpha$) X-ray diffraction patterns for milled samples using a Siemens D5000 set equipped with a position sensitive detector. All patterns were consistent with the samples being single phase with a B20 crystal structure. The lattice constant of the doped samples from the X-ray patterns depends linearly on the Mn, Co, and Al concentration[17,19-21], showing that Mn and Co successfully replace Fe and that Al replaces Si in the crystal structure. The stoichiometry of the samples was checked by performing energy-dispersive X-ray microanalysis (EDX) measurements which were all within error of nominal concentrations. The conductivity measurements were performed on rectangular samples cut by a string saw and polished with emery paper.

Thin Pt wires were attached to four linearly arranged silver paste contacts with an average spacing between voltage probes of 2 mm. Samples had an average cross sectional area of 1 x 0.5 mm$^2$. The resistivity ($\rho$) and magnetoconductance measurements were performed at 19 Hz using standard lock-in techniques in a dilution refrigerator with a 9 T superconducting magnet, and at higher temperature with a gas flow cryostat in a 5 T superconducting magnet. The magnetic susceptibility and magnetization of the same samples were measured in a Quantum Design SQUID magnetometer for temperatures between 1.7 and 300 K and fields between 0 and 5 T. Specific Heat measurements were performed using a standard semi-adiabatic heat pulse technique.

1. Al'tshuler, B.L., Aronov, A.G., Gershenson, M.E. & Sharvin, Yu.V. Quantum effects in disordered metal films. *Sov. Sci. Rev. A Phys.* **9**, 223-354 (1987).

2. Lee, P.A. & Ramakrishnan, T.V. Disordered electron systems. *Rev. Mod. Phys.* **57,** 287-337 (1985).

3. Cava, R. J. *et al*. Bulk Superconductivity at 91 K in a single-phase oxygen-deficient perovskite $Ba_2YCu_3O_{9-\delta}$. *Phys. Rev. Lett.* **58**, 1676-1679 (1987).

4. Kravchenko, S. V. *et al*. Electric field scaling at a B=0 metal-insulator transition in two dimensions. *Phys. Rev. Lett.* **77**, 4938-4941 (1996).

5. Custers, J. *et al*. The break-up of heavy electrons at a quantum critical point. *Nature* **424**, 524-527 (2003).

6. Mathur, N.D. *et al*. Magnetically mediated superconductivity in heavy fermion compounds. *Nature* **394**, 39-43 (1998).

7. Si, Q.M., Rabello, S., Ingersent, K. & Smith, J.L. Locally critical quantum phase transitions in strongly correlated metals. *Nature* **413**, 804-808 (2001).

8. Schroder, A. *et al*. Onset of antiferromagnetism in heavy-fermion metals. *Nature* **407**, 351-355 (2000).

9. Coleman, P. & Pepin, C. Singular Fermi liquid behavior in the underscreened Kondo model. *Phys. Rev. B*, **68** 220405 (1-4) (2003).

10. Mehta, P. *et al.* Regular and singular Fermi-liquid fixed points in quantum impurity models. *Phys. Rev. B* **72**, 104430 (1-10) (2005).

11. Posazhennikova, A. & Colman, P. Anomalous conductance of a spin-1 quantum dot. *Phys. Rev. Lett.* **94**, 036802 (1-4) (2005).

12. Sacramento, P.D. & Schlottmann, P. Thermodynamics of the *n*-channel Kondo model for general *n* and impurity spin *S* in a magnetic field. *J. Phys.: Condens. Matter* **3**, 9687-9696 (1991).

13. Wernick, J.H., Wertheim, G.K. & Sherwood, R.C. Magnetic behavior of monosilicides of 3D-transition elements. *Mat. Res. Bull.* **7,** 1431-1441 (1972).

14. Schlesinger Z. *et al.* Unconventional Charge Gap formation in FeSi. *Phys. Rev. Lett.* **71**, 1748-1751 (1993).

15. Mattheiss, L.F. & Hamann, D.R. Band-structure and semiconducting properties of FeSi. *Phys. Rev. B* **47,** 13114-13119 (1993).

16. Beille, J., Voiron, J. & Roth, M. Long period helimagnetism in the cubic-B20 $Fe_{1-x}Co_xSi$ and $Co_{1-x}Mn_xSi$ alloys. Sol. St. Commun.**47,** 399-402 (1983).

17. DiTusa, J.F. *et al*., Metal-insulator transitions in the Kondo insulator FeSi and classic semiconductors are similar. *Phys. Rev. Lett.* **78,** 2831-2834 (1997).

18. Chernikov, M.A. *et al.* Low-temperature transport, optical, magnetic, and thermodynamic properties of $Fe_{1-x}Co_xSi$. *Phys. Rev. B* **56,** 1366-1375 (1997).

19. DiTusa, J.F. *et al.*, Heavy fermion metal Kondo insulator transition in $FeSi_{1-x}Al_x$. *Phys. Rev. B*. **58,** 10288-102301 (1998).

20. Manyala, N. *et al*. Magnetoresistance from quantum interference effects in ferromagnets. *Nature* **404,** 581-584 (2000).

21. Manyala, N. *et al*. Large anomalous Hall effect in a silicon-based magnetic semiconductor. *Nature Materials* **3**, 255-262 (2004).

22. *See e.g* Ashcroft, N.W. & Mermin, N.D. *Solid State Physics*, Saunders College, Philadelphia (1976).

23. Pfleiderer, C., Julian, S.R. & Lonzarich, G.G. Non-Fermi-liquid nature of the normal state of itinerant-electron ferromagnets. *Nature* **414**, 427-430 (2001).

24. Rosenbaum, T.F., *et al*., Metal-insulator transition in a doped semiconductor. *Phys. Rev. B*. **27,** 7509-7523 (1983).

25. Husmann, A., et al., Dynamical Signature of the Mott-Hubbard Transition in $Ni(S,Se)_2$. *Science* **274**, 1874-1876 (1996).

26. von Molnar, S., Briggs, A., Flouquet, J. & Remenyi, G. Electron localization in a magnetic semiconductor: $Gd_{3-x}v_xS_4$. *Phys. Rev. Lett*. **51**, 706 (1984).

27. Paalanen, M.A., Graebner, J.E., Bhatt, R.N. & Sachdev, S. Thermodynamic behavior near a metal-insulator transition. *Phys. Rev. Lett*. **61**, 597-600 (1988).

28. Lakner, M., von Lohneysen, H., Langenfeld, A. & Wolf, P. Localized magnetic-moments in Si-P near the metal-insulator-transition. *Phys. Rev. B* **50**, 17064-17073 (1994).

29. Bhatt, R.N. & Lee, P.A. Scaling studies of highly disordered spin-1/2 antiferromagnetic systems. *Phys. Rev. Lett*. **48**, 344-347 (1982).

30. Sarachik, M.P *et al*. Scaling behavior in the magnetization of insulating Si:P. *Phys. Rev. B* **34**, 387-390 (1986).

31. Ghosh, S., Rosenbaum, T. F., Aeppli, G. & Coppersmith, S.N. Entangled quantum state of magnetic dipoles. *Nature* **425**, 48-51 (2003).

32.  Aeppli, G. & Fisk, Z. Kondo insulators. *Comments Cond. Mat. Phys.* **16**, 155–165 (1992).

Supplementary Information is linked to the online version of the paper at www.nature.com/nature.

We thank Zachary Fisk for discussions. JFD acknowledges support from the National Science Foundation, GA from a Wolfson-Royal Society Research Merit Award, and the Basic Technologies Program of the U.K. Research Councils.
14

**FIGURES**

**Figure 1** | Magnetic Susceptibility and Phase diagram: **a,** $\chi(T)$ for a subset of our $Fe_{1-x}Mn_xSi$ and $FeSi_{1-z}Al_z$ samples at $H=1$ kG with symbols identified in the figure. Lines are fits to a model that includes the sum of a Pauli susceptibility, a Curie Weiss term, and a thermally activated susceptibility previously used to describe pure FeSi. **b,** Curie constant, $C=g^2s(s+1)n$ vs. $x$, $y$, and $z$ from fits of the magnetic susceptibility, $\chi$, data to a Curie-Weiss form, $C/(T-\Theta_W)$. The solid and dashed lines are the Curie constants for $g=2$ and $S=1/2$, $S=1$, and $S=3/2$ moments where we have fixed $n$ equal to the Mn (or Co, Al) dopant densities. **c,** Weiss temperature ($\Theta_W$). Blue line is a linear fit to the $\Theta_W$ data for Co and Mn doped FeSi. Black line is a guide to the eye for the $FeSi_{1-z}Al_z$ data. Inset, $\Theta_W$ for $Fe_{1-x}Mn_xSi$ over the range $0<x<1$ showing the change to a positive $\Theta_W$ at $x>0.8$. **d,** Low temperature conductivity ($\sigma_o$) vs. nominal Mn ($x$), Co ($y$), and Al ($z$) concentrations ($H = 0$, 9 T). The solid lines represent fits to $\sigma_0 = \sigma_1(x/x_c - 1)^\nu$ with best fits corresponding to, $\sigma_1 = 1500 \pm 200$ $(460 \pm 100)$ $\Omega^{-1}cm^{-1}$, $x_c = 0.028 \pm 0.003$ $(0.014 \pm 0.003)$, and $\nu = 0.66 \pm 0.15$ $(0.7 \pm 0.1)$ in zero field (9 T) for $Fe_{1-x}Mn_xSi$, $\sigma_1 = 600 \pm 150$, $x_c = 0.007 \pm 0.003$, and $\nu = 0.85 \pm 0.1$ for $FeSi_{1-z}Al_z$, and $\sigma_1 = 1200 \pm 100$, $x_c = 0.018 \pm 0.005$, and $\nu = 0.37 \pm 0.1$ for $Fe_{1-y}Co_ySi$. For frames c-d data are represented by asterisks taken from Ref. 18. Error bars in frames b and c represent standard errors based on a least-squared fitting of data as in frame a. $FeSi_{1-z}Al_z$ data in frames a and d reproduced from Ref. [17] and [19].

**Figure 2** | Magnetotransport: **a**, Conductivity ($\sigma$) vs. temperature ($T$) of a subset of our Fe$_{1-x}$Mn$_x$Si samples at magnetic fields ($H$) identified in the figure. **b,** Color contour plot of the temperature and $x$ dependence of the conductivity of Fe$_{1-x}$Mn$_x$Si at $H = 0$ parameterized by the logarithmic derivative of $\sigma$ with respect to $T$, $\alpha = d\,ln(\sigma-\sigma_0)\,/\,dT$. The parameter $\sigma_0$ is chosen by fitting the data below 0.7 K to a power law form $\sigma=\sigma_0+\sigma_1 T^\alpha$. $\alpha = 0.5$ describes classic disordered metals, such as heavily doped semiconductors. **c,** $\alpha$, as in frame b, with $H = 9$ T. **d,** Color contour plot of the field and temperature dependence of $\alpha$ for the $x = 0.02$ sample.

**Figure 3** | Specific Heat: **a,** Specific heat divided by temperature ($C(T)\,/\,T$) vs. temperature for two Fe$_{1-x}$Mn$_x$Si samples at magnetic fields indicated in the figure. **b,** The change in entropy from its zero field value ($S(H) - S(0)\,/\,xR\ln 2$, where $R$ is the universal gas constant) calculated by integrating the $C(T)/T$ data in frame a after subtraction of the phonon term in the specific heat for $x=0.015$ (solid lines) and 0.03 (dashed lines). Grey (grey + green) shaded regions emphasize the increase in entropy at 15 K over the zero field data for the $x=0.03$ (0.015) sample. This entropy must be transferred from $T$'s below the lowest $T$ of our measurements at $H=0$ and brought into our $T$ window via the magnetic field (see text for details).

**Figure 4** | Underscreened Kondo effect: **a** compares underscreened[9-12], compensated and conventional ('overscreened') Kondo problems. **b-d** illustrate spin screening mechanisms in a magnetic semiconductor such as Fe$_{1-x}$Mn$_x$Si: Green arrows represent the Mn impurity spins (1) while red arrows represent the hole spin degrees of freedom (spin ½). **b,** At high temperatures there is a random distribution of rapidly fluctuating

impurity and hole spins. **c,** At temperatures below the Kondo temperature ($T_K$) the hole gas incompletely screens the impurity spins. **d,** At lower temperatures ($T < J$) many but not all of the impurity moments have been coupled to combinations, denoted by blue groupings, with singlet ground states. A few impurity moments are incompletely screened (yellow grouping) and contribute to entropy and are responsible for anomalous low temperature electrical properties.

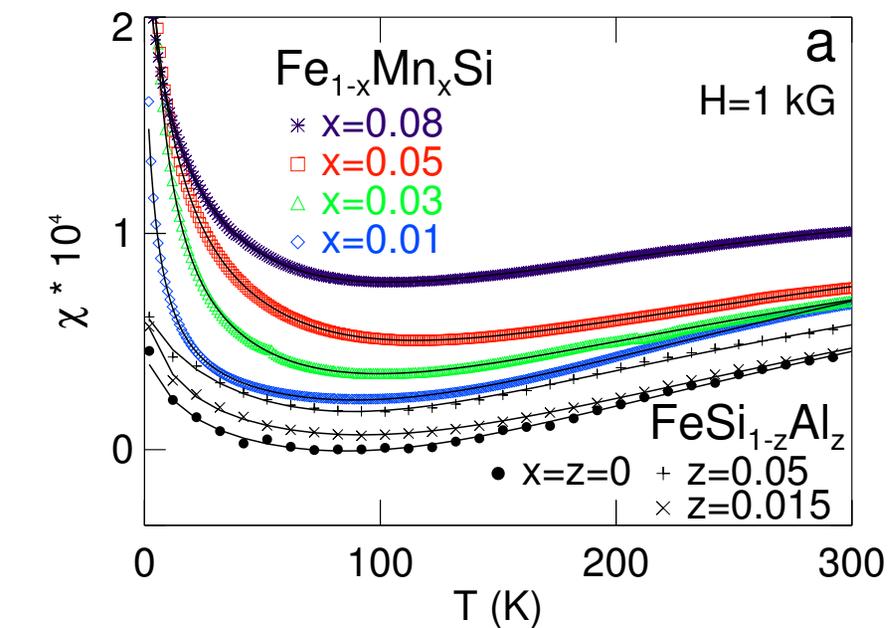
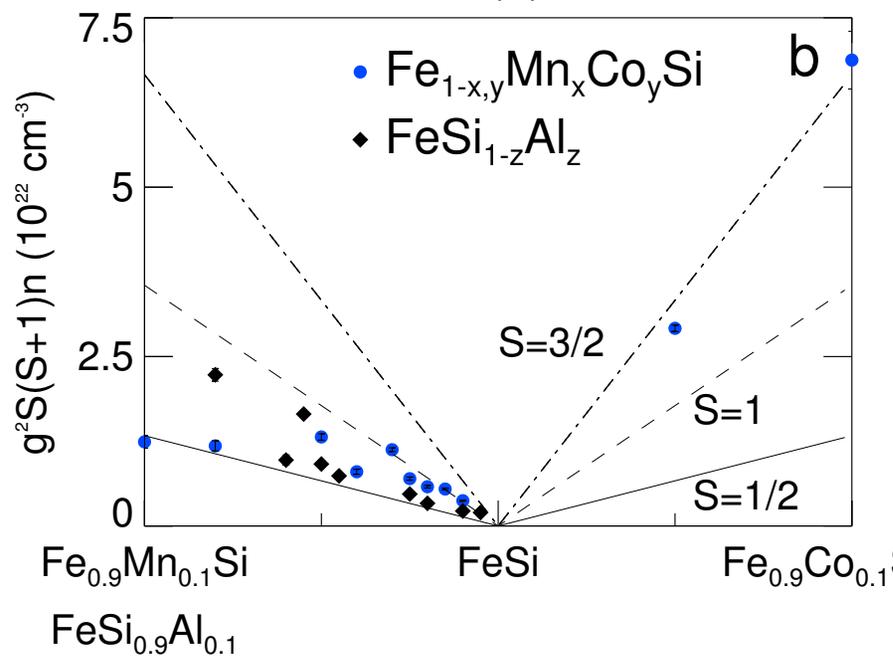
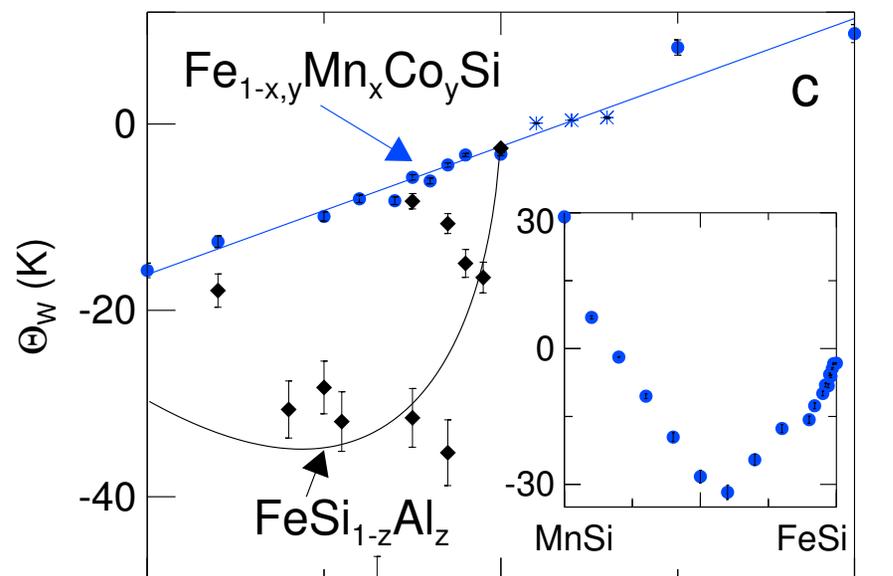
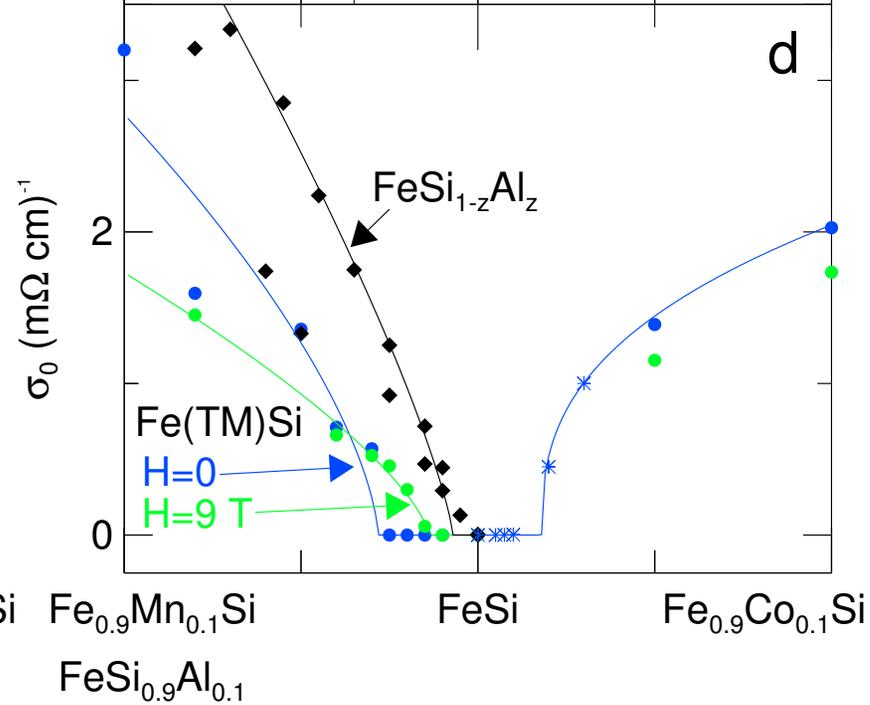

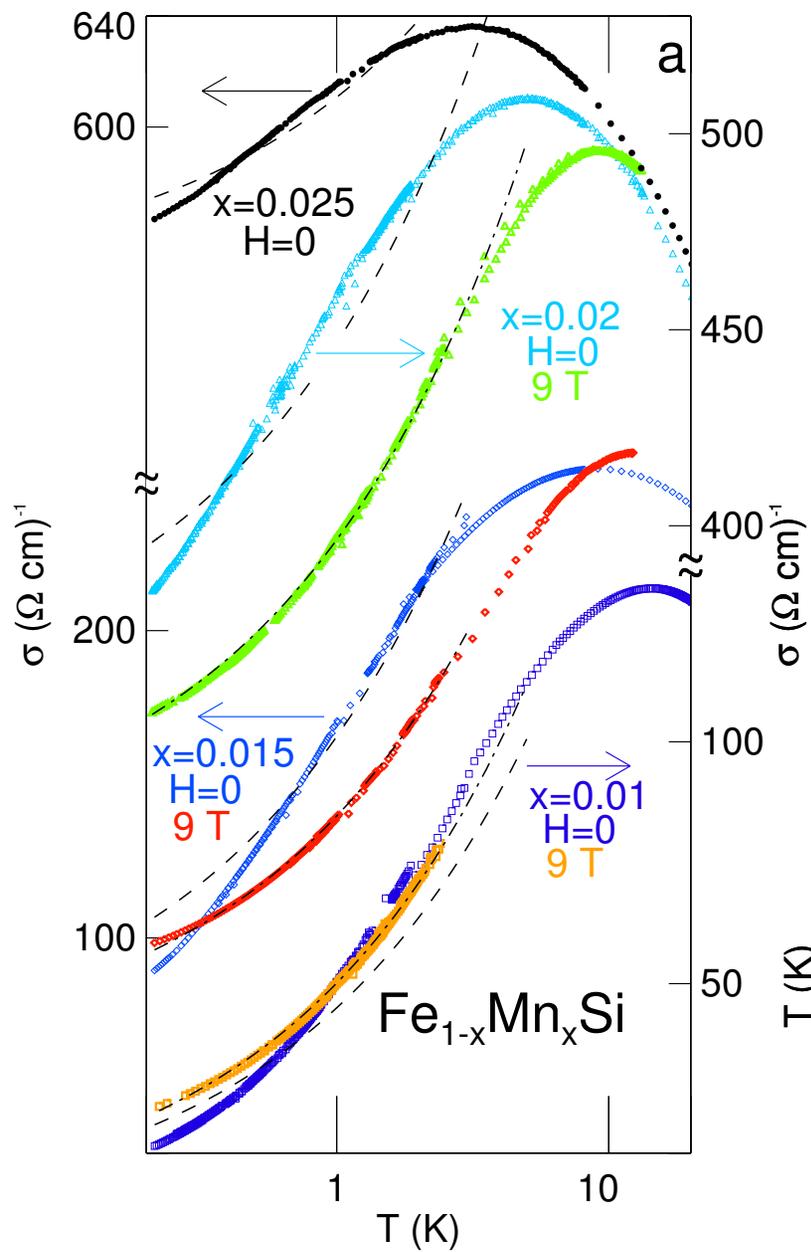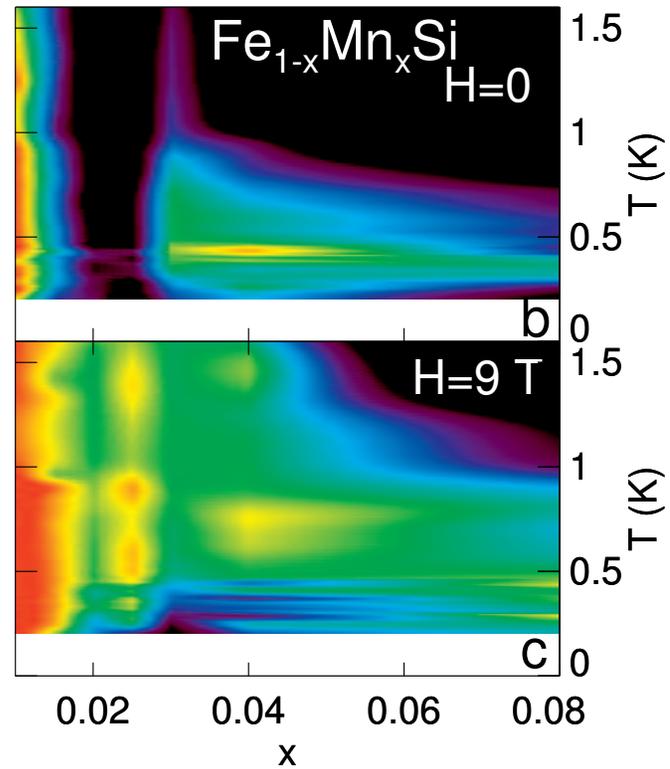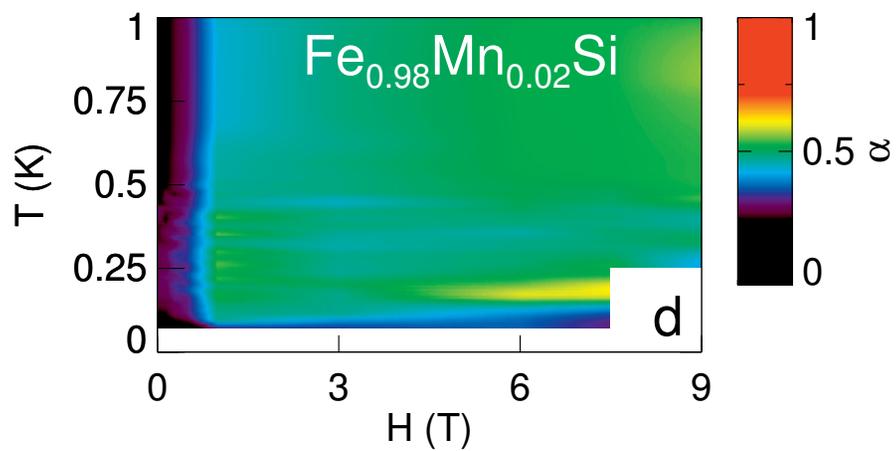

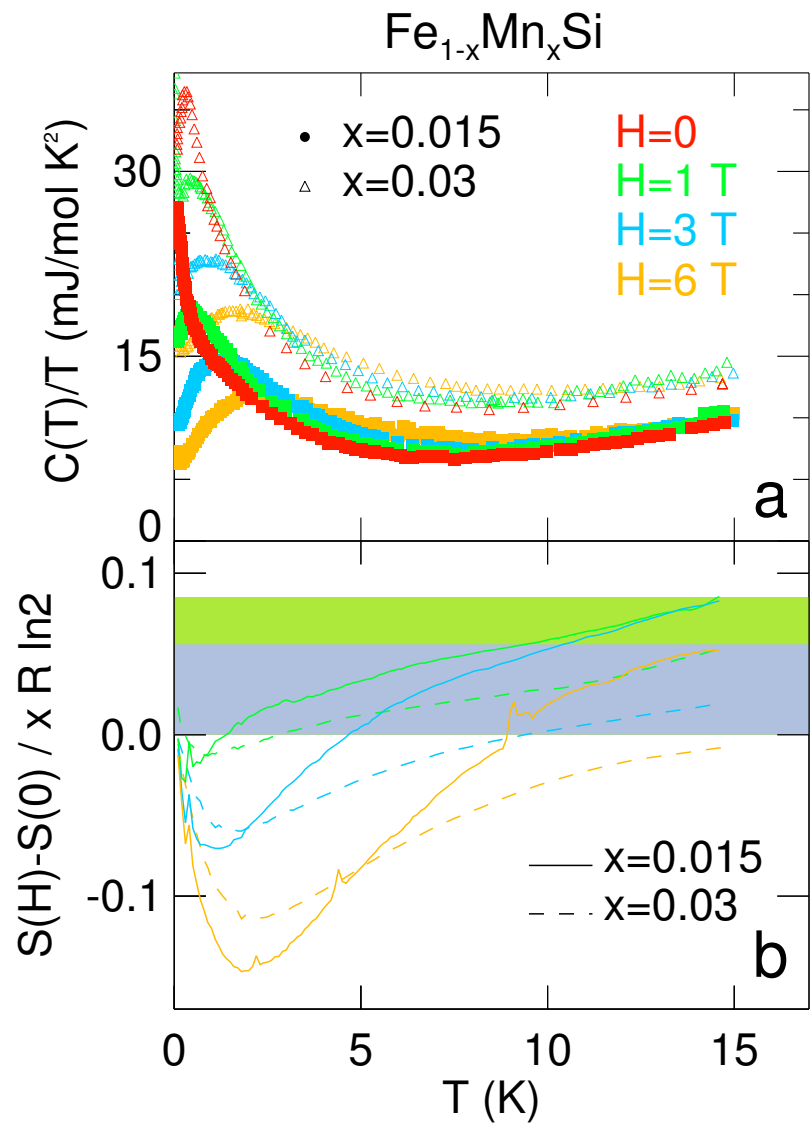

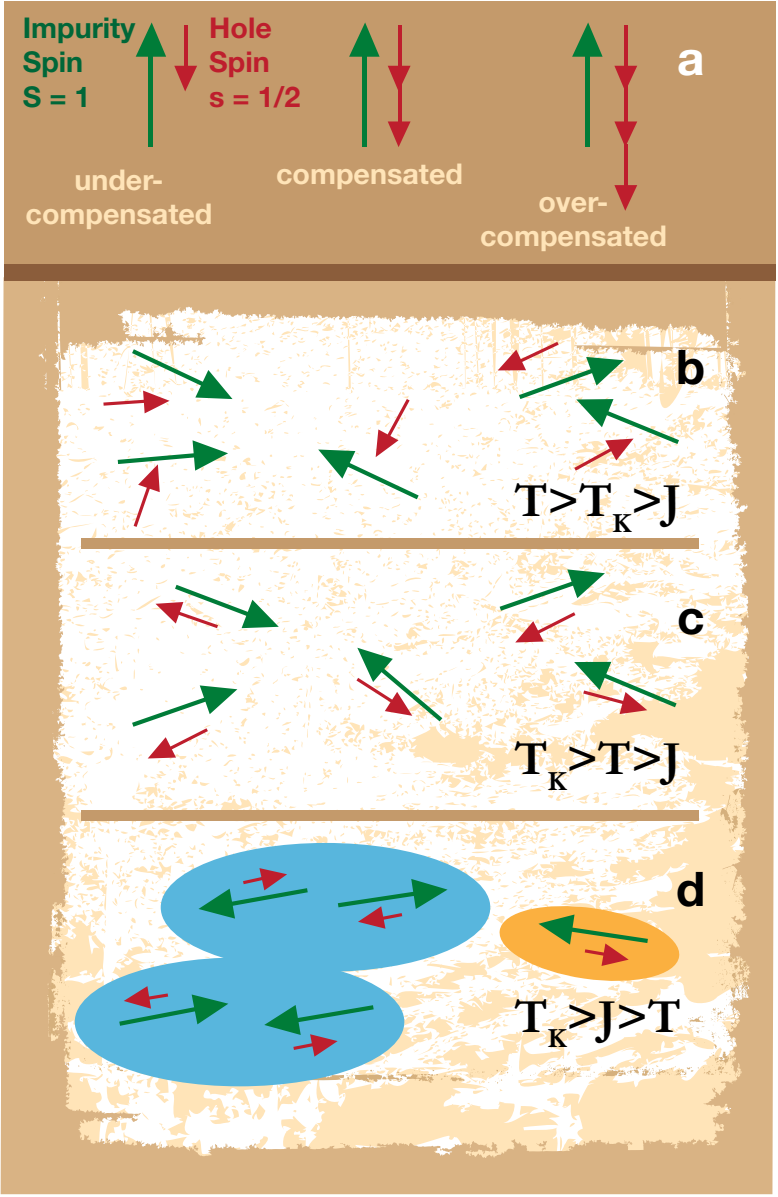

**Supplementary Information:**

Supplementary notes:

Magnetoresistance:

The magnetic field dependence of $\sigma$, shown in Fig. S1 for two Mn concentrations, is unlike that of standard disordered conductors in the same range of $x$, $T$, and $H$ where $\sigma(T)$ is unusual. At very low $T$'s a positive magnetoconductivity (MC) emerges in $Fe_{1-x}Mn_x$ Si at low fields that is not seen in the Al doped FeSi[1,2]. At high fields $\sigma(H)$ approaches an $H^{1/2}$ dependence of a similar magnitude to the $FeSi_{1-z}Al_z$ samples, so that again the application of magnetic field seems to restore the more standard behaviour. We note that the magnitude of the positive MC is larger than that predicted by weak localization theory and attempts to fit the data by the sum of theoretical expressions including e-e interactions and weak localization effects failed to reproduce the MC for any value of the parameters[3]. Both of these effects, a positive contribution to the MC, and a weakened temperature dependence of $\sigma$ (see Fig. 2c) cannot be understood within the standard e-e interaction picture that is so well documented for classic semiconductors[4], even when including the possible effects of weak localization.

Magnetization:

A comparison of the magnetic susceptibility of Al and Mn doped FeSi, along with a quantitative analysis of these data, reveals that Mn impurities take on a high spin state, close to $S=1$, that is poorly screened even at low temperatures. Observation of a similarly high spin state has also been previously reported for Gd when doped into the small gap semiconductor $SmB_6$[5]. The lack of screening of the moments by each other and the mobile holes is also apparent in the $H$-dependent magnetization curves. Fig. S2 shows $M(H)$ for three samples at 2 K and with $x$ near $x_c$ along with $M$ measured for a similar level of Al doping[2]. It is apparent in the figure that the Mn doped samples have larger $M$'s, and that the $H$ dependence does not correspond to the simple form (see Fig. S2b), where free moments compensate each other by forming singlets, that describes FeSi:Al and Si:P near their IM transitions[6-8]. However, if we add such a form, where the probability distribution of AFM coupling energies is characterized by a decaying power-law, $P(J)=J^{-\alpha}$, to a standard Brillouin function (corresponding to a delta function $\delta(J)$ contribution to $P(J)$) for the magnetization we obtain an acceptable description of the data for FeSi:Mn[6,9]. The key parameters for the uncoupled moments are the moment density, which is simply the high field saturation value $M_2$ of the Brillouin function and their size $S$, which we simply fix at the values given by the Curie-Weiss analysis of $\chi$ and recorded in Fig. 1b. The inset of Fig. S2a shows the upshot of our fits, namely that $M_2$ is largest just beyond the IM transition. In other words, in addition to antiferromagnetic interactions among the moments, there are also some local moments that do not participate in singlet formation via pairing with other local moments, and are therefore left to produce underscreened Kondo physics and the associated non-Fermi liquid (with disorder and interaction corrections) properties observed via our (extending to much lower temperature) electrical and specific heat measurements.

1. DiTusa, J.F. *et al.*, Metal-insulator transitions in the Kondo insulator FeSi and classic semiconductors are similar. *Phys. Rev. Lett.* **78,** 2831-2834 (1997).

2. DiTusa, J.F. *et al.*, Heavy fermion metal Kondo insulator transition in $FeSi_{1-x}Al_x$. *Phys. Rev. B*. **58,** 10288-102301 (1998).

3. Lee, P.A. & Ramakrishnan, T.V. Disordered electron systems. *Rev. Mod. Phys.* **57,** 287-337 (1985).

4. Rosenbaum, T.F., *et al.*, Metal-insulator transition in a doped semiconductor. *Phys. Rev. B*. **27,** 7509-7523 (1983).

5. Wiese, G., Schaffer, H., & Elschner, B. Possible $4f^7 5d^1$ ground state of Gd impurities in the mixed-valence compound $SmB_6$, observed with electron spin resonance. *Europhys. Lett*. **11**, 791-796 (1990).

6. Bhatt, R.N. & Lee, P.A. Scaling studies of highly disordered spin-1/2 antiferromagnetic systems. *Phys. Rev. Lett*. **48**, 344-347 (1982).

7. Sarachik, M.P *et al*. Scaling behavior in the magnetization of insulating Si:P. *Phys. Rev. B* **34**, 387-390 (1986).

8. Paalanen, M.A., Graebner, J.E., Bhatt, R.N. & Sachdev, S. Thermodynamic behavior near a metal-insulator transition. *Phys. Rev. Lett*. **61**, 597-600 (1988).

9. *See e.g* Ashcroft, N.W. & Mermin, N.D. *Solid State Physics*, Saunders College, Philadelphia (1976).

**SUPPLEMENTARY FIGURES**

**Figure S1** | Magnetoconductivity: Magnetoconductivity ($\Delta\sigma$) vs $H^{1/2}$ of a subset of our $Fe_{1-x}Mn_xSi$ and $FeSi_{1-z}Al_z$ samples at $T$s identified in the figure. $FeSi_{1-z}Al_z$ data reproduced from Ref. [1] and [2].

**Figure S2** | Magnetization: **a**, Magnetization ($M$) of 3 Mn doped FeSi samples along with $M$ of a similarly doped $FeSi_{1-z}Al_z$ sample at 2 K as a function of magnetic field ($H$). Dashed lines represent the best fit to a model of antiferromagnetically coupled spin pairs[6,7]. Dash-dotted line is the magnetization of paramagnetic (uncoupled) spin 1's suggested by a fit to the low field magnetic susceptibility for the $x$=0.025 sample. Solid lines are a model which includes both a population of AFM coupled impurity spins and a population of paramagnetic spins with a g-factor set equal to 2, a magnetic moment taken from the Curie constants shown in Fig. 1b of the paper, and density of paramagnetic spins as a free parameter. Shaded regions represent the contribution to $M(H)$ from the population of these free paramagnetic spins ($M_2$), with $x$=0.015 (light green), 0.025 (red), and 0.04 (gray). Inset, moment density ($M_2$) resulting from the fits vs. $x$. Error bars represent standard errors based on a least-squared fit of the data in the main part of frame a. **b** Difference, $\delta M$, between $M$s shown in a, and the best fit to the model of antiferromagnetically coupled spin pairs[6,7]. Symbols identified in a.

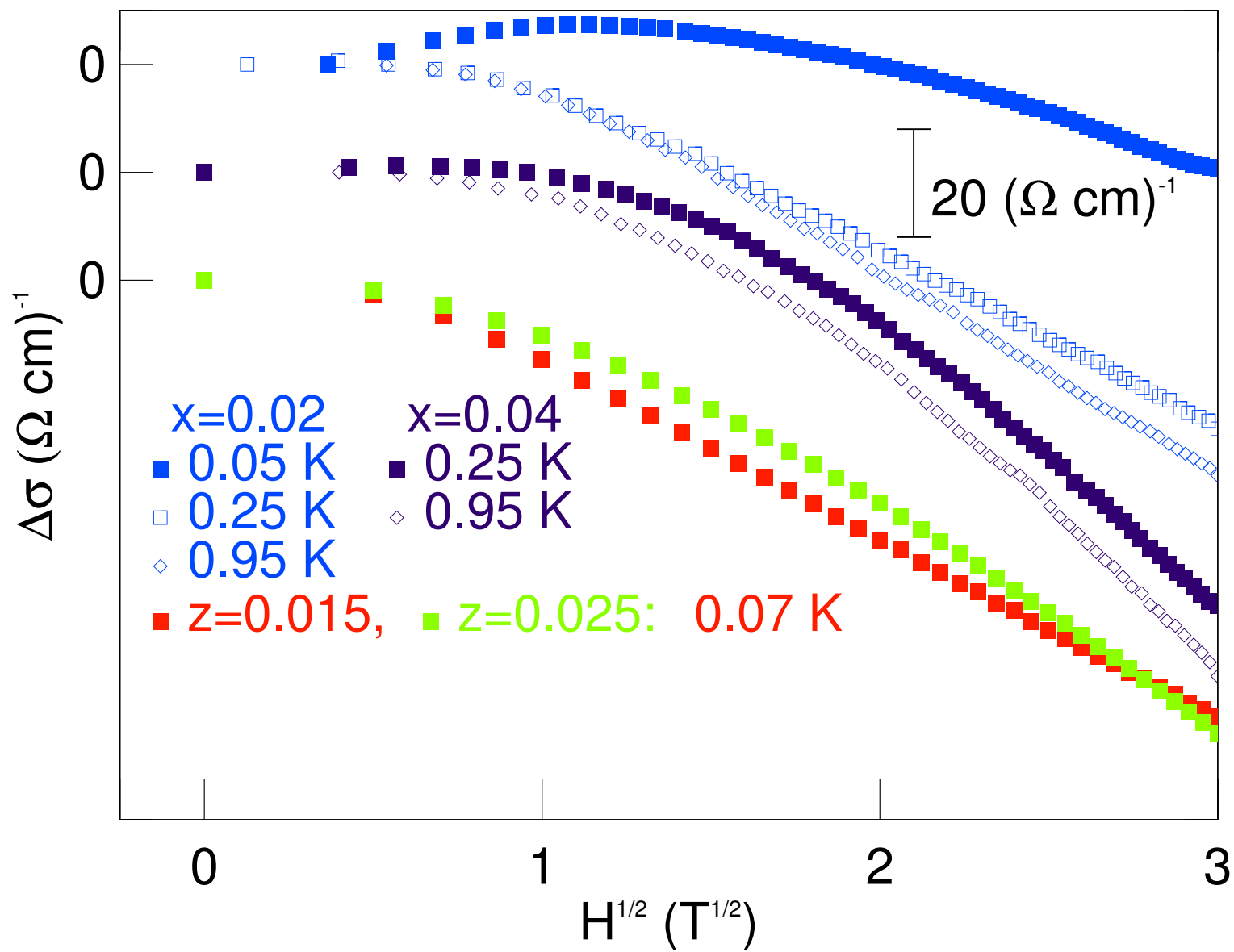

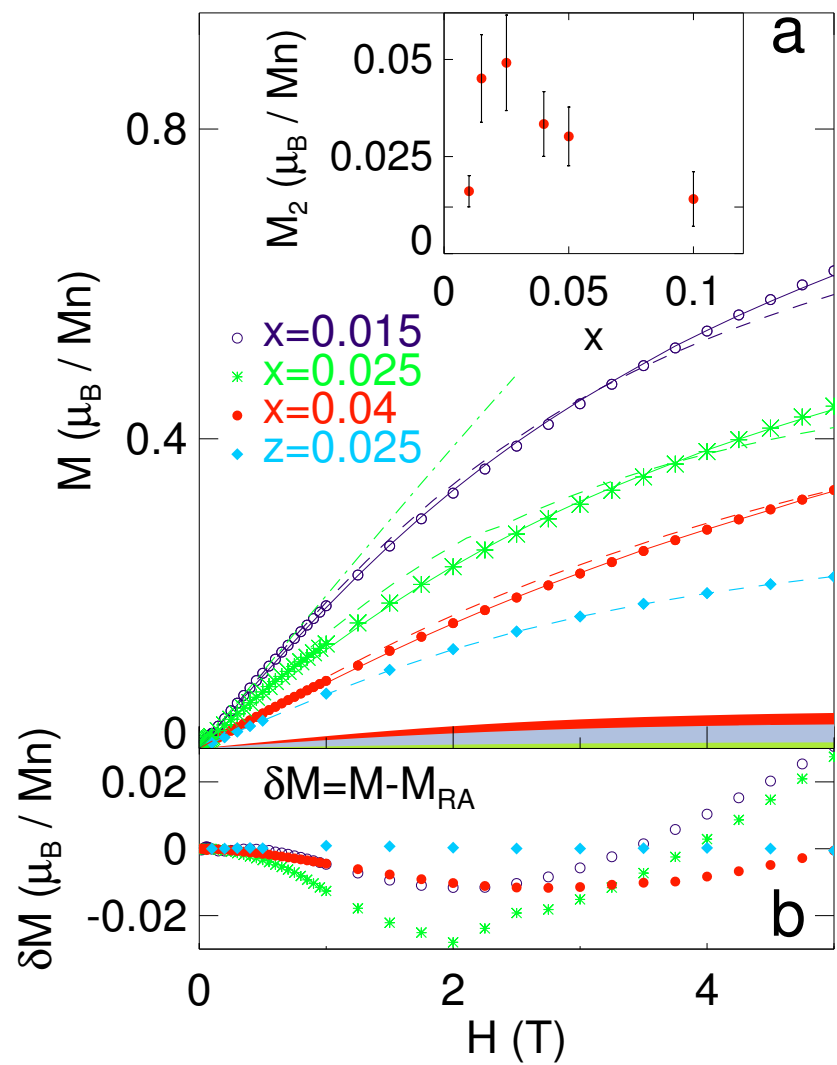